\documentclass[]{bio_arxiv}
\usepackage[colorlinks=true, urlcolor=citecolor, linkcolor=citecolor, citecolor=citecolor]{hyperref}

\usepackage{tabularx,colortbl}

\usepackage{graphicx}
\usepackage{epstopdf}
\graphicspath{{/Users/morganthompson/UCSF_files/Paper_Plots/}}
\usepackage{amsmath}

\numberwithin{equation}{section}

\newcommand{\rr}{\raggedright}

\begin{document}

\title{Comparing Evolutionary Rates Using An Exact Test for $2\times2$ Tables with Continuous Cell Entries}

\author{A. MORGAN THOMPSON\\[2pt]
\textit{Department of Bioengineering and Therapeutic Sciences, University of California San Francisco, San Francisco, CA  94158, USA}
\\[4pt]
M. CYRUS MAHER\\[2pt]
\textit{Department of Epidemiology and Biostatistics, University of California San Francisco, San Francisco, CA  94158, USA}
\\[4pt]
LAWRENCE H. URICCHIO\\[2pt]
\textit{Joint Graduate Group in Bioengineering, University of California Berkeley and San Francisco, San Francisco, CA  94158, USA}
\\[4pt]
ZACHARY A. SZPIECH\\[2pt]
\textit{Department of Bioengineering and Therapeutic Sciences, University of California San Francisco, San Francisco, CA  94158, USA}
\\[4pt]
RYAN D. HERNANDEZ$^\ast$\\[2pt]
\textit{Department of Bioengineering and Therapeutic Sciences, Institute for Human Genetics, Institute for Quantitative Biosciences (QB3), University of California San Francisco, San Francisco, CA  94158, USA}
\\[2pt]
{ryan.hernandez@ucsf.edu}}

\markboth%
{A. M. Thompson and others}
{Continuous Fisher's Exact Test}

\maketitle

\footnotetext{To whom correspondence should be addressed.}

\newpage
\begin{abstract}
{Assessing the statistical significance of an observed $2\times2$ contingency table can easily be accomplished using Fisher's exact test (FET). However, if the cell entries are continuous or represent values inferred from a continuous parametric model, then FET cannot be applied.  Such tables arise frequently in areas of biostatistical research including population genetics and evolutionary genomics, where cell entries are estimated by computational methods and result in cell entries drawn from the non-negative real line $\mathbb{R}^+_0$.  Simply rounding cell entries to conform to the assumptions of FET is an ill-suited approach that we show creates problems related to both type-I and type-II errors.  Pearson's $\chi^2$ test for independence, while technically applicable, is not often effective for these tables, as the test has several limiting assumptions that make application of this method inadvisable in many common instances (particularly with small cell entries). Here we develop a novel method for tables with continuous entries, which we term continuous Fisher's Exact Test (cFET).   Through simulations, we show that cFET has a close-to-uniform distribution of $p$-values under the null hypothesis of independence, and more power when applied to tables where the null hypothesis is false (compared to FET applied to rounded cell entries).  We apply cFET to an example from comparative genomics to confirm an overall increased evolutionary rate among primates compared to rodents, and identify several genes that show particularly elevated evolutionary rates in primates.  Some of these genes exhibit signatures of continued positive selection along the human lineage since our divergence with chimpanzee 5-7 million years ago, as well as ongoing selection in modern humans.}{FET, $2\times2$ contingency tables, tests for independence, categorical data analysis}
\end{abstract}

\section{Introduction}

Several statistical methods allow testing of the relationships between the cell entries of a $2 \times 2$ contingency table.  Pearson's $\chi^2$ is a popular test for independence between the rows and columns of a data table, and is calculated as follows:
\begin{displaymath}
\chi^2 \; = \; \sum_{i = 1}^r \sum_{j = 1}^c \frac{(O_{i,j} - E_{i,j})^2}{E_{i,j}}
\end{displaymath}
where $O_{i,j}$ and $E_{i,j}$ are the observed and expected values for the cell entries, respectively, $r$ the number of rows, and $c$ the number of columns.  For this test, $p$-values are obtained from the $\chi^2$ distribution with the appropriate degrees of freedom, and the null hypothesis is rejected when $p$ is smaller than a chosen critical point.  However, this test has several limiting assumptions.  Sample size is expected to be sufficiently large, or Type II error rates may increase.  In addition, cell entries for a $2\times2$ table are expected to be at least 5, below which Type I error rates tend to increase.  There is a correction available if the latter assumption proves false. \emph{Yates's Correction} (\citealp{yates}) subtracts 0.5 from the difference between observed and expected values for each cell entry, thereby increasing $p$-values.  However, the method tends to over-correct, leading again to increased Type II error rates.  

Another method is R.A. Fisher's eponymous Fisher's Exact Test (FET), which can handle low cell entries, and additionally benefits from the elimination of nuisance parameters by conditioning on sufficient statistics (see \citet{Agresti} for a review of the benefits of exact tests).  Fisher demonstrated that conditional on the row and column margin totals, the distribution of a cell entry follows the hypergeometric distribution.  The test is very useful for categorical data, with the caveat that it can only be used on contingency tables with discrete cell entries, since the hypergeometric distribution is a discrete probability distribution.

This limitation can be significant in many instances.  For example, contingency tables are sometimes constructed as averages (due to limited sampling ability), or represent estimates from continuous approximations.  In population genetics, it is common to ask whether species evolve at different rates, or whether natural selection has been operating on a given gene.  Data to address these questions can be summarized using $2 \times 2$ contingency tables in the following manner.  After sequencing and aligning a gene across multiple species, observed mutations can be cross-classified as either non-synonymous mutations (which occur when a DNA mutation causes a change in the amino acid sequence of a protein) or synonymous (which occur when a DNA mutation in a protein-coding region of the genome does not affect the amino acid sequence), and whether they occurred in one set of species or another.  

However, not all mutations can necessarily be observed, particularly for distantly related species.  This is because multiple DNA substitutions can serially arise at a single site, thereby masking the true number of events that have occurred.  The most common way to model the evolutionary process is to assume a continuous approximation, or to calculate the expected number of DNA substitutions from a Poisson process.  Clearly, both of these modeling approaches lead to non-integer estimates for substitution counts.  Moreover, for some common population genetics tasks, some cell entries are continuous while others are discrete. This arises for instance with the McDonald-Kreitman test  (\citealp{mcd}), which compares substitutions across species (continuous) versus polymorphic mutations within a species (often discrete).   

To address challenges such as these, we derive a statistical test that extends FET to the non-negative real line $R^+_0$.  We call our procedure the continuous Fisher's Exact Test (cFET).  We evaluate its performance through simulations under the null hypothesis, and find that when all cell entries are treated as continuous, cFET has a distribution of p-values that is nearly uniform (as compared to FET, which is highly skewed and conservative).  When the null hypothesis is false, we find that cFET is a more powerful test than the common practice of rounding cell entries to their nearest integer and applying FET.  We then apply cFET to a genome-wide data set of protein coding genes that have been aligned across human, rhesus macaque, mouse and rat to examine differences in evolutionary rates between primates and rodents.  We find several statistically significant genes that highlight the interesting evolutionary history of primates.

\section{Methods}
\subsection{Mixture of discrete and continuous cell entries}
\label{sec2.1}

In this section, we introduce a framework for extending the Binomial distribution to allow for a mixture of continuous and discrete values.  In the next section, we extend the model further to consider the general case of two continuous values.  We then derive a new exact test for $2 \times 2$ contingency tables with discrete and/or continuous cell entries.

Consider a general $2 \times 2$ contingency table as shown in Table \ref{genTable}, where $A$ and $B$ are the primary random variables of interest, with the additional random variables $R_1$, $R_2$, and $C_1$ representing the two row sums and first column sum (respectively).  Following the derivation of FET, assume an observed table with $A=a, B=b, R_1=r_1, R_2=r_2$, and $C_1=c_1$, where $a, b, r_1, r_2, c_1 \in \mathbb{Z}$.  Further assume that the distribution of $A|R_1 \sim \text{Bin}(R_1, p_1)$, and $B|R_2 \sim \text{Bin}(R_2, p_2)$.  This could arise, for example, if $A\sim \text{Poisson}(p_1R_1)$ and $B\sim \text{Poisson}(p_2R_2)$.  However, if $a, b \in \mathbb{Z}$ but $r_1, r_2 \in \mathbb{R}_0^+$ (with $a\le r_1$ and $b \le r_2$), then the strict binomial modeling assumption is violated.  Instead, we can extend the Binomial distribution using Newton's Generalized Binomial Theorem, which states the following result for $x \in \mathbb{R}$ (or even $x \in \mathbb{C}$):
\begin{equation}
(\alpha+\beta)^x = \sum_{i=0}^\infty \binom{x}{i}\alpha^{i}\beta^{x-i}
\end{equation}
where $\binom{x}{i} = \frac{1}{i!}\prod_{j=0}^{i-1}(x-j)$.  For the general scenario shown in Table \ref{genTable} with a mixture of continuous and discrete cell entries, we say that the conditional distribution of $A$ given $R_1$ follows a continuous/discrete Binomial distribution, which we write as $A | R_1 \sim cd\text{Bin}(R_1, p_1)$.  We define this distribution as follows
\begin{equation}
P(A=a | R_1, p_1) = \kappa_{R_1,p_1}\binom{R_1}{a}p_1^a(1-p_1)^{R_1-a}
\end{equation}
\noindent where $\kappa_{R_1,p_1}$ is a normalizing constant.  The corresponding probability for $B$ could also be constructed.  Drawing from the analogy of $cd\text{Bin}()$ to the Binomial distribution, we note that if $A \sim cd\text{Bin}(R_1, p)$ and $B \sim cd\text{Bin}(R_2, p)$, then $C_1=A+B \sim cd\text{Bin}(N, p)$, where $N=R_1+R_2$.  The proof of this is analogous to proving that the sum of Binomial random variables (with the same probability of success) is also a Binomial random variable by writing 
\begin{equation}
P(A+B = c | R_1+R_2, p) = \sum_{i=0}^rP(A=i|R_1, p)P(B=c-i|R_2,p)
\end{equation}
and utilizing Newton's Generalized Binomial Theorem to show that $\sum_{i=0}^r\binom{R_1}{i}\binom{R_2}{c-i} = \binom{R_1+R_2}{c}$. We next turn to the case of all four cell entries being continuous before deriving our new statistical test.

\subsection{Continuous cell entries}
\label{sec2.2}
Now suppose we have Table \ref{genTable} where $A=a, B=b, R_1=r_1, R_2=r_2, C_1=c_1 \in \mathbb{R}_0^+$.  In this case, all cell entries are continuous, and there is yet another extension of the binomial model that can applied.  The Gamma function is an extension of the factorial function to the real line $\mathbb{R}$, shifted by one.  That is, for a natural number $n$, $\Gamma(n+1) = n!$, and for any $\alpha \in \mathbb{R}$, $\Gamma(\alpha) \; = \; \displaystyle\int_0^{\infty}t^{\alpha - 1}e^{-t}\,dt$.  This function has been used to extend the Binomial Coefficient to the non-negative real line $\mathbb{R}^+_0$ (\citealp{Fowler}).  We will refer to this quantity as the continuous Binomial Coefficient, and define it as follows for $N, x \in \mathbb{R}^+_0$ with $N \ge x$:
\begin{equation}
\binom Nx \; := \; \frac{\Gamma(N + 1)}{\Gamma(x + 1)\Gamma(N - x + 1)}
\label{M1}
\end{equation}

\noindent In this way, if $N$ and $x$ are non-negative integers, then (\ref{M1}) returns the classic Binomial Coefficient.  If $N \in \mathbb{R}$ and $x \in \mathbb{Z}$, then (\ref{M1}) returns Newton's Generalized Binomial coefficient from Section \ref{sec2.1}.  Otherwise, this provides the functional form necessary to interpolate.  Note that $\binom{N}{x}$ is similar, but not equal to the inverse Beta function.  We can now use this to extend the Binomial probability distribution to the non-negative real line, where if $X=x$ and $N \in \mathbb{R}^+_0$ and $X \sim c\text{Bin}(N,p)$, then
\begin{equation} 
f(X = x | N,p) = \kappa_{N,p}\,\binom Nx p^x(1 - p)^{N - x} I_{x \in [0, N]}
\label{M2}
\end{equation}
\noindent where $\kappa_{N,p} = \left(\displaystyle\int_0^N \binom Nv p^v (1 - p)^{N - v}\,dv\right)^{-1}$ is a normalizing constant.  A closed-form solution for $\kappa_{N,p}$ has not been found, but it is easily obtained numerically and, as we show below, is irrelevant for the development of our exact test.  Returning to Table \ref{genTable}, assume that $A|R_1 \sim c\text{Bin}(R_1, p_1)$ and $B|R_2 \sim c\text{Bin}(R_2, p_2)$.  Similar to section \ref{sec2.1}, drawing from the analogy of $c\text{Bin}()$ to the Binomial distribution, we note that $C_1=A+B \sim c\text{Bin}(N, p)$, where $N=R_1+R_2$.

\begin{figure}[h]
\centering\includegraphics[width=\textwidth]{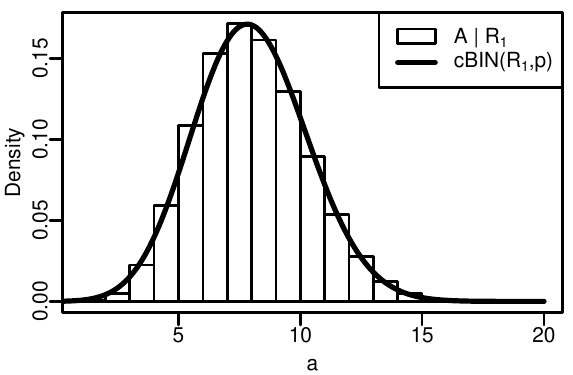}
\caption{Comparison of the cBin(24, 1/3) distribution derived in Section \ref{sec2.2} versus the distribution of a $A|R_1$, where $R_1=A+X$, with $A\sim\Gamma(8,1)$, $X\sim\Gamma(16,1)$, and $A+X \in (23.75, 24.25)$.}
\label{cBinVgamma}
\end{figure}

We note that if $A\sim \Gamma(\alpha_1, 1)$ and $X \sim \Gamma(\alpha_2, 1)$, then while $Z=A/(A+X) \sim \beta(\alpha_1, \alpha_2)$, the distribution of $A|A+X$ does not follow a known distribution. However, Figure \ref{cBinVgamma} shows that cBin() is a good approximation, at least for the parameters evaluated here.  This is an important development, as the Gamma distribution is a natural extension of the Poisson distribution to $\mathbb{R}^+$.  For Figure \ref{cBinVgamma}, we simulated $10^7$ $\{A,X\}$ pairs with $A\sim \Gamma(8, 1)$ and $X \sim \Gamma(16, 1)$.  We then plot the distribution of $A$ conditional on $R=A+X$ being within a defined range.  Explicitly, we plot the distribution of $\{A=a | R=r \in (23.75, 24.25)\}$, and see very good agreement when we overlay the cBin(24, 1/3) distribution.

\subsection{Deriving cFET}
 \label{sec2.3}
Our goal is to test the null hypothesis $H_0 : p_1 = p_2$ for both the case of a mixture of discrete and continuous cell entries presented in section \ref{sec2.1} and for the case in which all four cell entries are continuous as presented in section \ref{sec2.2}.  To do so, we test the conditional distribution of $A$ given the row ($R_1, R_2$) and column ($C_1$) sums, analogous to what is done in FET.  We derive our exact test for Table \ref{genTable} assuming $A=a, B=b, R_1=r_1, R_2=r_2, C_1=c_1 \in \mathbb{R}_0^+$.  The derivation for the case of $A=a, B=b, C_1=c_1 \in \mathbb{Z}$ but $R_1=r_1, R_2=r_2 \in \mathbb{R}_0^+$ is analogous, exchanging integration for summation.

Under the null hypothesis in FET, $p_1 = p_2$, and the distribution of $A|R_1, R_2, C_1 \sim \text{Hypergeometric}(R_1, R_2, C_1)$.  The following algebraic manipulations lead us to a continuous version of the hypergeometric distribution:
\begin{align} 
f(A = a | R_1 = r_1, R_2 = r_2, C_1 = c_1, p) \; &= \; \frac{f(A = a, A + B = c_1 | R_1 = r_1, R_2 = r_2, p)}{f(C_1 = c_1|R_1 = r_1, R_2 = r_2, p)} \nonumber \\
&\nonumber\\
&= \; \frac{f(A = a|R_1 = r_1, p)f(B = c_1 - a|R_2 = r_2, p)}{f(C_1 = c_1|R_1 + R_2 = r_1 + r_2, p)} \nonumber \\
&\nonumber\\
&= \; \frac{\kappa_{r_1, p} \, \binom{r_1}a p^a(1 - p)^{r_1 - a}\,\, \kappa_{r_2, p} \, \binom{r_2}{c_1 - a} p^{c_1 - a}(1 - p)^{r_2 - (c_1 - a)}}{\kappa_{r_1 + r_2} \, \binom{r_1 + r_2}{c_1} p^{c_1}(1 - p)^{r_1 + r_2 - c_1}}\nonumber\\
&\nonumber\\
&= \; \frac{\kappa_{r_1,p}\kappa_{r_2,p}}{\kappa_{r_1 + r_2, p}} \cdot \frac{\binom{r_1}a \, \binom{r_2}{c_1 - a}}{\binom{r_1 + r_2}{c_1}} \cdot \frac{p^{c_1}(1-p)^{r_1 + r_2 - c_1}}{p^{c_1}(1 - p)^{r_1 + r_2 - c_1}}\nonumber\\
&\nonumber\\
&= \; \kappa_{r_1, c_1, r_1+r_2} \frac{\binom{r_1}a \binom{r_2}{c_1 - a}}{\binom{r_1 + r_2}{c_1}} \nonumber \\
&\nonumber\\
&= \; \kappa_{r_1, c_1, n} \frac{\binom{r_1}a \binom{n - r_1}{c_1 - a}}{\binom n{c_1}}
\label{M3}
\end{align}

Since $\displaystyle\int f(x|R_1 = r_1, R_2 = r_2, C_1 = c_1, p)\,dx = 1$, we must have $\kappa_{r_1, c_1, n} = \left(\displaystyle\int{ \frac{\binom{r_1}x \binom{n - r_1}{c_1 - x}}{\binom n{c_1}}\, dx}\right)^{-1}$, demonstrating that the conditional distribution is independent of the nuisance parameter $p$.  We will refer to the resulting distribution as a continuous hypergeometric distribution, and write $A|R_1, C_1, N \sim c\text{Hypergeometric}(R_1, C_1, N).$

The test for significance is similar to that of FET, but using numerical integration.  Let $a_{\min} = \max(0, C_1 - R_2)$ and $a_{\max} = \min(R_1, C_1)$ be the minimal and maximal values $A$ can take on such that all the cell entries are non-negative when conditioning on the marginal totals $R_1, \:C_1$ and $N$.  Then for the general case of Table \ref{genTable}, and $f(a|r_1, r_2, c_1, n)$ in (\ref{M3}), let
\begin{equation}
\Psi \; = \; \{x : x \in (a_{\min}, \, a_{\max}), \; f(x | r_1, c_1, n) \leq f(a | r_1, r_2, c_1, n)\}
\label{M4}
\end{equation}
To get a p-value for $H_0$, we use
\begin{equation}
p \; = \; \int_{x \in \Psi} f(x|r_1, r_2, c_1, n)\, dx
\label{M5}
\end{equation}

\noindent For the case of a mixture of discrete and continuous cell entries presented in section \ref{sec2.1}, the test is directly analogous to FET, and can be similarly derived by replacing numerical integration with summation.

\subsection{cFET as an exact test for relative evolutionary rates}
In population genetics and comparative genomics, one is often interested in comparing orthologous genes across species to detect instances of natural selection (see \citealp{Maher.Hernandez}).  However, if species are sufficiently diverged, the number of DNA substitutions that have arisen cannot necessarily be observed.  This is because multiple substitutions may have occurred at the same site, thereby masking the true number of events.  Instead, researchers often infer the rate at which substitutions have occurred under a parametric model.  To detect natural selection, the goal is to estimate the rate of non-synonymous substitutions per non-synonymous site in a gene ($dN$) and the rate of synonymous substitutions per synonymous site in a gene ($dS$).  Since non-synonymous mutations change the makeup of the protein coded for by the gene, the $dN/dS$ ratio can be seen as an indicator of selective pressure acting on a protein-coding gene.  A $dN/dS$ ratio of one indicates no selection, a ratio greater than one implies positive Darwinian selection, and less than one implies purifying (stabilizing) selection.  

Another formulation of the test is to construct a $2\times 2$ contingency table as shown in Table \ref{relRateTable} for the example of comparing selective pressures in rodents versus primates.  In this case, we need the number of synonymous and non-synonymous substitutions that have occurred in each lineage, either within rodents (determined by comparing the mouse and rat genomes) or within primates (determined by comparing human and rhesus genomes).  Given an observed number of substitutions for each cell entry of this $2\times 2$ table, one could then perform a FET.  But, given the large divergence time between each of these sets of species, parsimoniously counting substitutions can be biased (see \citealp{Hernandez.ea_2007}).  By estimating the total number of positions in a gene at which a non-synonymous ($N$) or synonymous ($S$) mutation could arise, the expected number of non-synonymous and synonymous substitutions can then be obtained from the products $N\,\cdot\,dN$ and $S\,\cdot\,dS$.  But, since $dN$ and $dS$ are rates they are always continuous.  We therefore have $N_p, S_p \in \mathbb{R}_0^+$ instead of $\mathbb{Z}$ as the expected number of nonsynonymous and synonymous substitutions in primates (and analogously $N_r$ and $S_r$ for rodents).  Instead of FET, we can use cFET (derived in section \ref{sec2.3}) on the resulting $2 \times 2$ table with continuous cell entries.  This test can be interpreted as a relative evolutionary rates test, where the null hypothesis is that the odds ratio of the $2 \times 2$ table is equal to one (i.e., that the primate and rodent $dN/dS$ ratios are equal).

Another scenario that frequently arises in population genetics is where one column in a $2 \times 2$ table is more easily observed than the other.  In the McDonald-Kreitman test (\citealp{mcd}), one compares synonymous and non-synonymous variants that are either substitutions across species or polymorphic within a species.  Depending on the species being compared, an order of magnitude less polymorphisms might be observed compared to substitutions.  It is therefore often reliable to assume that there are not multiple mutations segregating at the same site within the population (i.e., the infinite sites assumptions, where polymorphisms can be observed directly).  The number of substitutions, however, can be large enough that statistical estimation of the expected number of substitutions is required as above.  In this case, the resulting $2 \times 2$ table consists of a column of polymorphisms that are natural numbers, and a column of substitutions that are non-negative real numbers.  Such a $2\times2$ table can be analyzed using cFET based on the distribution derived in section \ref{sec2.1}.  

\section{Results}
\subsection{Simulations}

To test the power and type-I error of cFET, we simulated $2\times2$ tables and generated graphical representations of comparisons between the application of cFET and FET (after rounding) to these tables.  Rather than drawing cell entries from \emph{c}Bin, they were simulated via the Gamma Distribution $\Gamma(\alpha, \beta)$, where $\alpha$ is the ``shape'' parameter and $\beta$ is the ``scale'' parameter.  The mean of this distribution is $\alpha\beta$ and its variance is $\alpha\beta^2.$  Here, we always set $\beta = 1$ so that the mean equaled the variance in our simulations, to mimic the properties of the Poisson distribution extended to $\mathbb{R}^+$.    We chose to simulate data from the Gamma distribution to better mimic the cell entries we might observe in data rather than drawing cell entries directly from the distributions we derive in sections \ref{sec2.1}-\ref{sec2.2}.

\begin{figure}[h]
\centering\includegraphics[width=\textwidth]{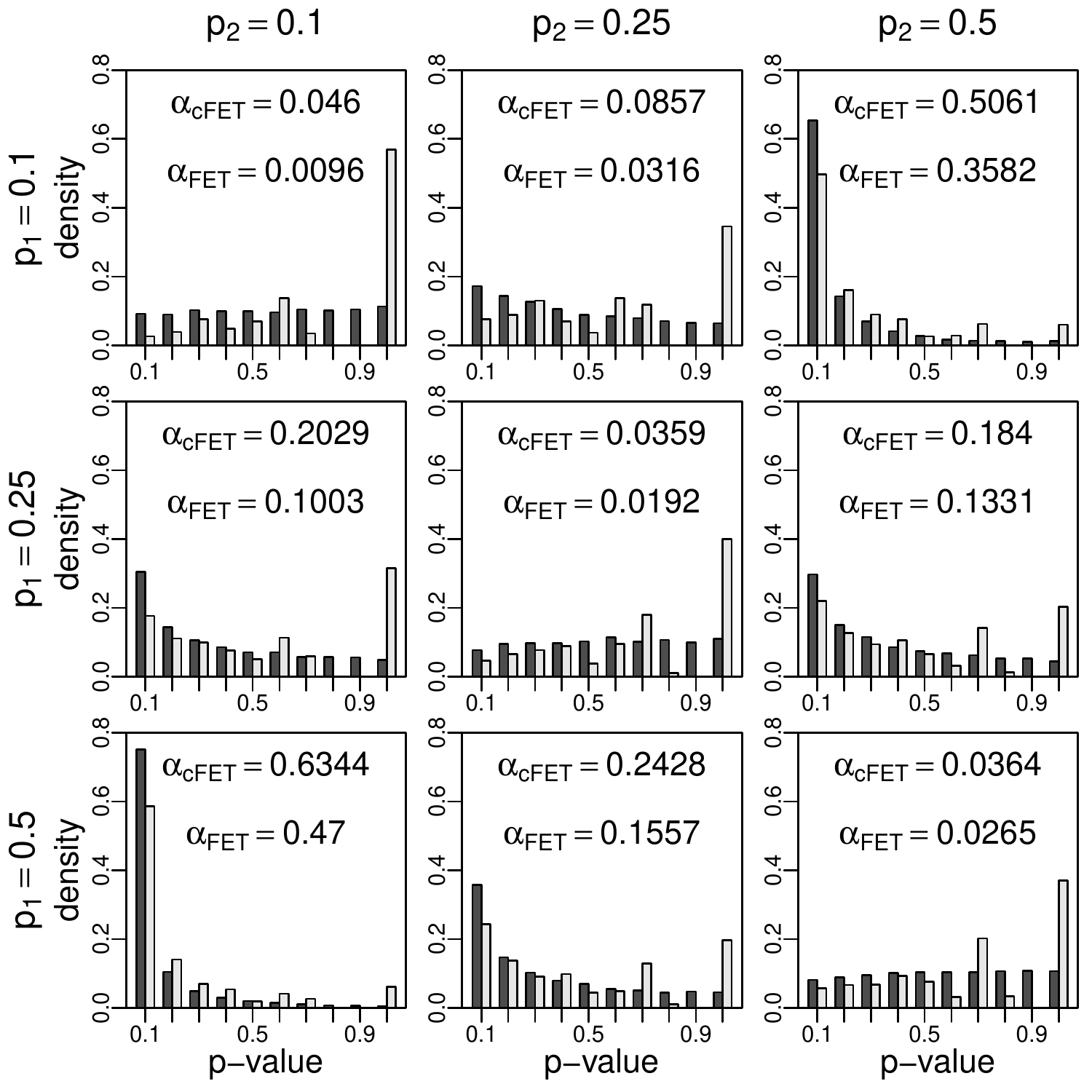}
\caption{Results of applying both cFET (dark bars) and the FET (light bars) method to 10,000 $2\times2$ tables in which cell entries are drawn independently from a $\Gamma(\alpha, \beta)$ distribution with $\mathbb{E}(R_1) = 8$ and $\mathbb{E}(R_2) = 16$ (see Table \ref{genTable}) for three different values of $p_1$ and $p_2$.  Note that cell entries were rounded to the nearest integer before applying FET.  cFET provides a more uniform distribution of $p$-values under the null hypothesis, and has increased power when the null hypothesis is false.}
\label{PvalDist}
\end{figure}

\begin{figure}[h]
\centering\includegraphics[width=\textwidth]{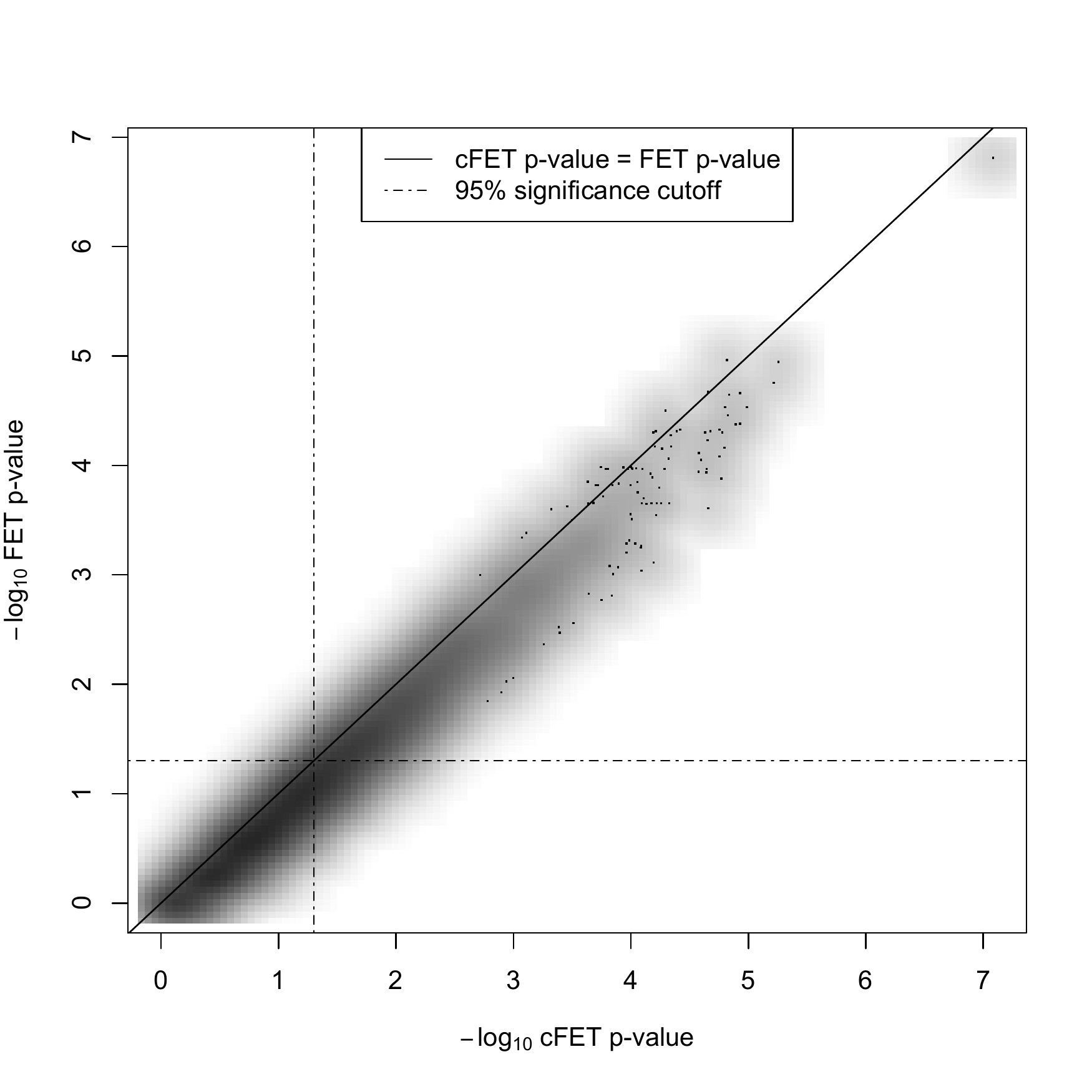}
\caption{A smoothed scatter plot of the $-\log_{10}$($p$-values) for FET versus cFET from 10,000 simulated $2\times2$ tables in which $p_1 = 0.1$ and $p_2 = 0.5$, and $\mathbb{E}(R_1)=\mathbb{E}(R_2)=10$.}
\label{PvalScat}
\end{figure}

Considering the setup in Table \ref{genTable}, let $X = R_1-A$ and $Y = R_2-B$.  We evaluated cFET with $\mathbb{E}(R_1) = 8$ and $\mathbb{E}(R_2) = 16$.  For each pair of $p_1$ and $p_2$ in $\{0.1, 0.25, 0.5\}$, we simulated 10,000 data tables with $A \sim \Gamma(p_1\mathbb{E}(R_1),1)$, $B \sim \Gamma(p_2\mathbb{E}(R_2),1)$, $X \sim \Gamma((1-p_1)\mathbb{E}(R_1),1)$, and $Y \sim \Gamma((1-p_2)\mathbb{E}(R_2),1)$.  For each simulated data table, we applied cFET, then rounded the cell entries to the nearest integer and applied FET.  The distribution of p-values for each test (as well as the proportion of simulated tables rejected at the 0.05 level) are reported in Figure \ref{PvalDist}.  Figure \ref{PvalScat} shows a smoothed scatter plot of the $-\log_{10}(p-\text{value})$ for FET (applied to rounded cell entries) versus cFET when $\mathbb{E}(R_1) = \mathbb{E}(R_2) = 10$, $p_1=0.1$, and $p_2=0.5$, along with vertical and horizontal lines to indicate $-\log_{10}(0.05)$.  Points in the upper right quadrant indicate tables for which both tests reject $H_0$.  In contrast, points in the upper left and lower right indicate tables in which only one of the methods (FET or cFET, respectively) rejected $H_0$.  The skew of $p$-values below the diagonal line indicates that in general, cFET has more power than applying FET after rounding.  For these parameters, cFET rejects $H_0$ for ${\sim}28\%$ more tables FET (applied to rounded cell entries). Tests evaluating cFET with a mixture of discrete and continuous cell entries resulted in $p$-value distributions that were nearly indistinguishable from FET, and are not shown.

\subsection{Faster evolutionary rates in primates compared to rodents}

We obtained 4,375 human protein coding genes that could be accurately aligned to rhesus macaque, mouse, and rat (\citealp{Kosiol}).  To test for differences in the evolutionary rate in the primate versus rodent lineage, we used PAML (\citealp{yang}) to infer the number of synonymous and non-synonymous DNA substitutions that occurred in the primate lineage (e.g., human or rhesus) or the rodent lineage (e.g., mouse or rat).   We created $2\times2$ tables following the structure shown in Table \ref{relRateTable}, and analyzed all tables using cFET.  

\begin{figure}[h]
\centering\includegraphics[width=\textwidth]{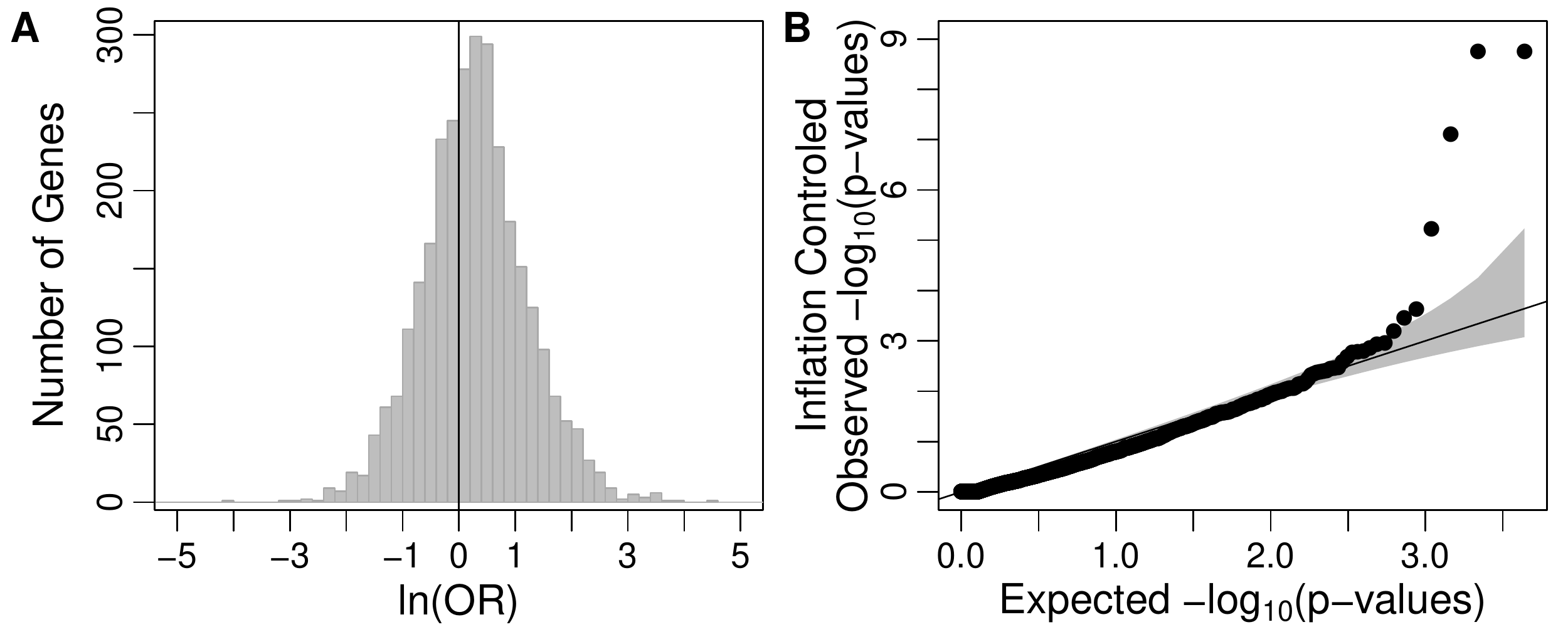}
\caption{Faster evolutionary rates in primates compared to rodents.  A)  The distribution of the ln(odds ratio) of genes as shown in Table \ref{relRateTable}.  Values greater than 0 suggest faster evolutionary rates in primates ($\approx58\%$ of genes).  B) QQ-plot of the log transformed $p$-values comparing evolutionary rates among primates vs rodents from cFET.  Genes with zero mutations received $p$-values of 1.  Inflation control was applied, resulting in an inflation factor $\lambda=1.79$.}
\label{ORQQ}
\end{figure}

Figure \ref{ORQQ}A shows the distribution of ln(Odds ratio)$= \ln\left(\frac{N_p S_r}{S_p N_r}\right)$ across all tables.  The interpretation of this quantity is the log fold change in $dN/dS$ between species.  Under the null hypothesis that rodents and primates evolve at the same rates, we would expect this distribution to be centered around zero.  Instead, we find that the distribution is shifted to toward positive values, with 520 genes ($11.9\%$) having an odds ratio $>1$ with $p<0.05$, as compared to only 127 genes ($2.9\%$) having an odds ratio $<1$ with $p<0.05$.  We find that the average evolutionary rate for primates has been greater than rodents for the genes we were able to accurately align.  This observation is consistent with previous observations using different methods (\citealp{Kosiol}).  

We next sought to statistically test the observed $2\times2$ tables using cFET.  In Figure \ref{ORQQ}B, we show the QQ-plot for our observed -log$_{10}$($p$-value) versus their expectations after controlling for global inflation of significant $p$-values.  We find that several genes significantly deviate from the null hypothesis.  The genes with $p<1\times10^{-5}$ surpass Bonferroni correction, and are shown in Table \ref{SigGenes}.  We find two genes LYG1 and LHB have $p$-values below the resolution of our implementation of cFET ($p<2\times10^{-16}$).  Consistent with the odds ratio distribution shown in Figure \ref{ORQQ}A, eight of our top nine genes show significantly elevated evolutionary rates in primates compared to rodents.  The alignments for all genes shown in Table \ref{SigGenes} have been manually verified, and confirmed using two independent protocols including MUSCLE (\citealp{Edgar}; downloaded from \hyperref[http://www.ncbi.nlm.nih.gov/homologene]{http://www.ncbi.nlm.nih.gov/homologene}) and MOSAIC (\citealp{Maher.Hernandez}).

To understand whether our top genes show continued rapid evolution in the human lineage, we performed a genome-wide analysis of McDonald-Kreitman tables, where substitutions were inferred to have occurred along the human lineage since our divergence with chimpanzee (using \citealp{Maher.Hernandez} alignments), and polymorphisms from the 1000 Genomes Project (\citealp{TGP}).  The continuous/discrete form of cFET failed to identify any genome-wide significant tables due to lack of power.  We instead used the Poisson random effects model SnIPRE (\citealp{Eilertson}).  Interestingly, UBXN10 exhibits one of the strongest signatures of human lineage selection in the genome, suggesting that much of this gene's signal of rapid evolution in primates may be due to human-lineage-specific evolutionary pressures.

We then inspected signatures of very recent natural selection in humans using the iHS statistic (\citealp{Voight}; \citealp{Szpiech.Hernandez}).  This statistic looks for abnormally long haplotypes at high frequency, indicative of recent positive selection.  Our top gene from the cFET analysis, LYG1, is within the top $10\%$ of all genes in the genome for this statistic.  This suggests that this gene may be under continued evolutionary pressure in modern humans.

\section{Discussion}

In the analysis of $2\times2$ contingency tables, exact tests, such as Fisher's Exact Test (FET), are appealing because conditioning on sufficient statistics causes nuisance parameters to cancel out.  However, FET cannot be applied to $2\times2$ tables with non-integer entries.  Such tables frequently arise, such as when some or all of the cell entries are not directly observed, but, rather, are inferred as the expectation from a model.  Here we present two such scenarios where this arises, one in population genetics (the McDonald-Kreitman test when outgroup species are sufficiently diverged) and one from comparative genomics (in the case of a relative evolutionary rates test).  

To surmount the problem of cell counts that lie along the non-negative real line, we developed cFET, a continuous analog of Fisher's Exact Test.  Our test is based on the extension of the notion of binomial coefficients to the real line using either Newton's Generalized Binomial Theorem (in the case of a mixture of discrete and continuous cell entries) or interpolating between integer values using the $\Gamma$ function.  We show that in the case of all cell entries being continuous, cFET has a distribution of $p$-values that are nearly uniform under the null hypothesis, and has more power than FET when applied to cell entries that have been rounded.  

We applied cFET to a genome-wide set of 4,375 human transcripts that could be accurately aligned across rhesus macaque, mouse and rat.  We find evidence for an overall increased evolutionary rate among primates compared to rodents.  This is consistent with previous observations using different analyses (\citealp{Kosiol}).  We find 13 genes with $p$-values that exceed the Bonferroni corrected threshold (see Table \ref{SigGenes} for the top 9 with $p<1\times10^{-5}$).  In contrast, applying FET to rounded cell entries yields eight genes that surpass the stringent Bonferroni correction.  This observation is consistent with our simulations showing increased power for cFET compared to FET.  These genes highlight fundamental aspects of mammalian evolution, eight of which exhibit significantly elevated rates among primates compared to rodents.  Our top gene from this analysis is lysozyme G-like 1 (LYG1), which is thought to be an anti-bacterial enzyme, and highlights the importance that pathogens have had in shaping the human genome.  Interestingly, LYG1 also shows signatures of very recent positive selection in modern humans (in the top $10\%$ of all human proteins).  This suggests that there are likely ongoing selective pressures operating on the human genome, perhaps the result of continued host-pathogen interactions.  Another gene, UBXN10 has not been well studied, but exhibits among the top scores in the entire proteome for natural selection along the human lineage since our divergence with chimpanzee 5-7 million years ago.  It remains to be determined what fraction of these genes that have undergone significantly increased evolutionary rates in primates compared to rodents are involved in shaping uniquely human phenotypes and disease (\citealp{Maher.Uricchio.ea}).

While we have largely explored applications of cFET within population genetics and comparative genomics, there are potentially many other uses of cFET.  For example, in gene expression studies, one often obtains quantitative expression values that are continuous.  To test whether a gene is over- or under-expressed in a particular tissue compared to one or more housekeeping genes, one could apply cFET as an exact test for differential expression.  The method developed here is flexible, and can readily be extended in multiple ways whenever observations lie outside of $\mathbb{Z}$.

\section{Software}

cFET is implemented in the \texttt{C} programming language, and is available upon request or from the corresponding author's webpage (\hyperref[http://bts.ucsf.edu/hernandez_lab]{http://bts.ucsf.edu/hernandez\_lab}).  Numerical integration has been implemented using Romberg's method as implemented in \citet{Press}. The current implementation provides a highly efficient implementation of the original Fisher's Exact Test, as well as both extensions described herein (for continuous as well as a mixture of discrete and continuous data).  Given a $2\times2$ table, cFET allows you to specify the desired test, or will automatically choose the proper test based on the table's contents.

\section*{Acknowledgments}
We would like to thank Carlos D. Bustamante for initially inspiring this research, and Nicolas Strauli for providing comments on the manuscript.  This work was partially supported by the National Institutes of Health (grant numbers P60MD006902, UL1RR024131, 1R21HG007233, 1R21CA178706, and 1R01HL117004) and an Alfred P. Sloan Foundation Research Fellowship to R.D.H.  M.C.M. was supported by the Epidemiology and Translational Science program at the University of California, San Francisco, a National Institutes of Health F31 Predoctoral Fellowship (grant number 1 F31 CA180609-01), and a University of California, San Francisco Lloyd M. Kozloff Fellowship.  L.H.U. was supported by an Achievement Rewards for College Scientists (ARCS) Fellowship.

{\it Conflict of Interest}: None declared.

\newpage

\begin{table}[h]	
	\resizebox{0.6\textwidth}{!}{\begin{minipage}{0.55\textwidth}
	\caption{A model of cross-classified data in a $2\times2$ table with margin totals given by Row Sum and Column Sum.  Class$_1$ and Class$_2$ are arbitrary classifications, with Class$_1^C$ and Class$_2^C$ the respective complements.  The primary random variables of interest are indicated by observations (depicted by lowercase letters).}\medskip
	\begin{tabular}{r | c | c | c l }
		\multicolumn{1}{c}{} & \multicolumn{1}{c}{$\text{Class}_2$} & \multicolumn{1}{c}{$\text{Class}_2^C$} & \multicolumn{1}{c}{Row Sum}\\
		\cline{2-3}
		$\text{Class}_1$ & $A=a$ & $X=R_1 - A$ & $R_1=r_1$ \\ \cline{2-3}
		$\text{Class}_1^C$ & $B=b$ & $Y=R_2 - B$ & $R_2=r_2$ \\ \cline{2-3}
		\multicolumn{1}{c}{Column Sum}& \multicolumn{1}{c}{$C_1=c_1$} & \multicolumn{1}{c}{$N - C_1$} & \multicolumn{1}{c}{$N=n$} \\ 
	\end{tabular}
	\label{genTable}
	\end{minipage} }
\end{table}

\begin{table}[h]
	\resizebox{0.6\textwidth}{!}{\begin{minipage}{0.55\textwidth}
	\caption{A $2\times2$ contingency table showing the number of nonsynonymous and synonymous substitutions in primates versus rodents.}\medskip
	\begin{center}
	\begin{tabular}{r | c | c | }
		\multicolumn{1}{c}{} & \multicolumn{1}{c}{Nonsynonymous} & \multicolumn{1}{c}{Synonymous}  \\ \cline{2-3}
		\text{Primate} & $\text{N}_p$ & $\text{S}_p$  \\ \cline{2-3}
		\text{Rodent} & $\text{N}_r$ & $\text{S}_r$  \\  \cline{2-3}
	\end{tabular}
	\label{relRateTable}
	\end{center}
	\end{minipage} }
\end{table}

\begin{center}
	\begin{table}[h]\footnotesize
	\caption{Description of top 9 genes with significant tables according to cFET Method.} \medskip
	\begin{center}
	\begin{tabular}[l]{| m{0.6cm} | m{1.1cm} | m{2cm} | m{.9cm} | m{1.1cm} | m{.6cm}| m{0.6cm} |m{6cm} |}
		\hline
		Rank & Gene & Cell Entries$^1$ & Length &  cFET$^2$ & MK$^3$ & iHS$^4$ & Gene Description \\ \hline
		\rr 1 & LYG1 & \begin{tabular}{c | c}\textbf{15.8} & \textbf{5.2} \\ \hline 0.0 & 24.9 \\  \end{tabular}   & 1015bp &  \textbf{$<$2e-16} & 0.69 & \textbf{0.91} & Belongs to the glycosyl hydrolase 23 family.  Found in extracellular region and may participate in catabolic, metabolic and peptidoglycan catabolic processes. (\citealp{irwin}) \\ \hline
		\rr 2 & LHB & \begin{tabular}{c | c}\textbf{12.8} & \textbf{0.0} \\ \hline 0.0 & 13.2\\  \end{tabular}  & 426bp & \textbf{$<$2e-16} & 0.13 & NA & Encodes the beta subunit of luteinizing hormone (LH). LH promotes spermatogenesis and ovluation. \\ \hline
		\rr 3 & ADCY10 & \begin{tabular}{c | c}\textbf{65.8} & \textbf{19.7} \\ \hline 55.6 & 131.1\\  \end{tabular}  & 5018bp &  \textbf{1.8e-13} & 0.82 & 0.11 &  Thought to function as a general bicarbonate sensor throughout the body, and also may play an important role in generation of cAMP in spermatozoa. \\ \hline
		\rr 4 & TSPAN8 & \begin{tabular}{c | c}\textbf{51.4} & \textbf{9.9} \\ \hline 23.9 & 51.3\\ \end{tabular}  &  1104bp & \textbf{4.3e-10} & 0.40 & 0.29 &  Cell-surface glycoprotein that helps regulate cell development, activation, growth and motility.\\ \hline
		\rr 5 & PTCD3 & \begin{tabular}{c | c} \textbf{75.4} & \textbf{37.3} \\ \hline 28.9 & 61.5 \\  \end{tabular} & 1940bp & \textbf{6.3e-7} & 0.16 & 0.42 & This protein associates with the mitochondrial small ribosome subunit and regulates translation. (\citealp{davies}) \\ \hline
		\rr 6 & DNAJB8 & \begin{tabular}{c | c}\textbf{7.0} & \textbf{0.0} \\ \hline 3.0 & 37.0 \\ \end{tabular}  & 1082bp & \textbf{1.9e-6} & 0.12 & 0.44 & Belongs to the DNAJ/HSP40 protein family that regulate chaperone activity.  Suppresses aggregation and toxicity of polyglutamine proteins.  \\ \hline
		\rr 7 & MZB1 & \begin{tabular}{c | c}4.0 & 12.5 \\ \hline \textbf{15.5} & \textbf{0.2} \\  \end{tabular} & 826bp & \textbf{5.1e-6} & 0.82 & NA & Promotes immunoglobulin M assembly and secretion.  Acts as a molecular chaperone or as an oxidoreductase.  May be involved in regulation of apoptosis. \\ \hline
		\rr 8 & TRIM42 & \begin{tabular}{c | c} \textbf{27.7} & \textbf{33.8} \\ \hline 6.8 & 59.0 \\  \end{tabular} & 2540bp & \textbf{7.6e-6} & 0.55 & 0.49 & A member of the tripartite motif (TRIM) family, which includes 3 zinc-binding domains and a coiled-coil region. \\ \hline
		\rr 9 & UBXN10 & \begin{tabular}{c | c} \textbf{6.9} & \textbf{2.4} \\ \hline 5.5 & 60.2 \\  \end{tabular} & 1335bp & \textbf{9.6e-6} & \textbf{0.99} & NA & The UBX domain-containing proteins constitute the largest family of known Cdc48/p97 cofactors.  Cdc48/p97 belongs to the highly-conserved, essential, chaperone-related protein family of AAA ATPases (\citealp{schuberth}). \\ \hline
	\end{tabular}
	\end{center}
	\label{SigGenes}
	\footnotesize{\noindent$^1$ See Table \ref{relRateTable} for format.  Row with highest $dN/dS$ ratio is in bold.\\
	$^2$ $p$-value obtained from cFET.\\
	$^3$ Genome-wide percentile for McDonald-Kreitman test for ancient human-lineage selection.  Genes in the top 10-percentile are in bold.\\
	$^4$ Genome-wide percentile for iHS.  Genes in the top 10-percentile are in bold.}
	\end{table}
\end{center}

\newpage

\end{document}